\documentclass[doublecol]{epl2} 

\usepackage{textcomp}
\usepackage{multirow}

\title{Estimation of number of runaway electrons per avalanche in Earth's atmosphere}
\shorttitle{Number of runaway electrons per avalanche  in Earth's atmosphere} 

\author{
	T. Khamitov\inst{1,2}\thanks{E-mail: \email{timuruh@mail.ru}} 
	\and 
	A. Nozik\inst{1,2}\thanks{E-mail: \email{nozik.aa@mipt.ru}} \and 
	E. Stadnichuk\inst{1,2}\thanks{E-mail: \email{egrstadnichuk@yandex.ru}} \and 
	E. Svechnikova\inst{3}\thanks{E-mail: \email{svechnikova@ipfran.ru}}  \and
	M. Zelenyi\inst{1,2}\thanks{E-mail: \email{mihail.zelenyy@phystech.edu}}
}

\shortauthor{M. Zelenyi \etal}

\institute{                 
	\inst{1} Moscow Institute of Physics and Technology (National Research University) - 1 “A” Kerchenskaya st., Moscow, 117303, Russian Federation \\
	\inst{2} Institute for Nuclear Research of RAS - prospekt 60-letiya Oktyabrya 7a, Moscow 117312\\
	\inst{3} Institute of Applied Physics of RAS - 46 Ul'yanov str., 603950, Nizhny Novgorod, Russia
}

\pacs{52.90.+z}{Other topics in physics of plasmas and electric discharges}
\pacs{02.70.Uu}{Applications of Monte Carlo methods}
\pacs{13.40.-f}{Electromagnetic processes and properties}

\begin{document}
	\abstract{
		The connection between thunderstorms and relativistic runaway electron avalanches is an important topic that has attracted the attention of many researchers. Among other things, there are a lot of various simulations of the dynamics of electron avalanches. This article was written mostly in response to the article "The critical avalanche of runaway electrons" by Evgeny Oreshkin et al, which shows rather large numbers for an estimate of the number of runaway electrons, but it also contains the results of our own simulation and comparison with other papers.
	}
	
	\maketitle
	
	\section{Introduction}
	The phenomenon of relativistic runaway electron avalanche (RREA) development in an electric field is a generally accepted mechanism for enhancement of low energy gamma~($10\un{keV}-100\un{MeV}$) and electron fluxes in the Earth's atmosphere \cite{Dwyer2014,CHILINGARIAN201468}. This mechanism draws additional attention because it could shed some light on the problem of generation of lightning discharge which up to this point does not have a satisfactory explanation. Moreover, a number of observed phenomena (like Terrestrial Gamma-Ray Flashes~(TGFs) \cite{Dwyer2012} and Thunderstorm Ground Enhancements~(TGE) \cite{CHILINGARIAN201468}) is believed to have an origin related to RREA. However, certain conditions of the formation of avalanches and the dynamics of its development remain unknown. Review of  articles presenting an analytical models of avalanche development can be found in~\cite{Dwyer2012}. 
	Numerical modeling of energetic particle propagation in the air leading to an avalanche generation is presented in another set of papers \cite{moss2006, DwyerSmith2005, skeltved2014}.
	
	A recent paper \cite{Oreshkin_2018} presents the results of the 3D Monte-Carlo simulation of runaway avalanche development. The paper is focused on the avalanche-to-streamer transition. One of the main results of \cite{Oreshkin_2018} is the estimation of the number of electrons in one avalanche: $10^{17}$ - $10^{18}$. The value is significantly larger than values presented in other simulations~\cite{Gurevich:2001, dwyer2003fundamental,dwyer2011low}, which show about $10^6$ fast electrons and $10^{10}$ slow electrons per avalanche in normal conditions (air pressure, gas composition, etc.). The difference could be attributed to some non-standard approximations made in the article.
	
	We present the results of RREA simulation carried out using a software designed by our group on the basis of GEANT4 libraries. The simulation results show the doubtfulness of conclusions derived in~\cite{Oreshkin_2018}, where the number of electrons in RREA is highly overestimated. In order to illustrate a physical origin of limitation of the number of particles, we provide a simple analytical upper bound estimation.
	
	\section{Analytical estimation of the number of runaway electrons}
	
\begin{figure}[h]
		\centering
		\includegraphics[width=0.45\textwidth]{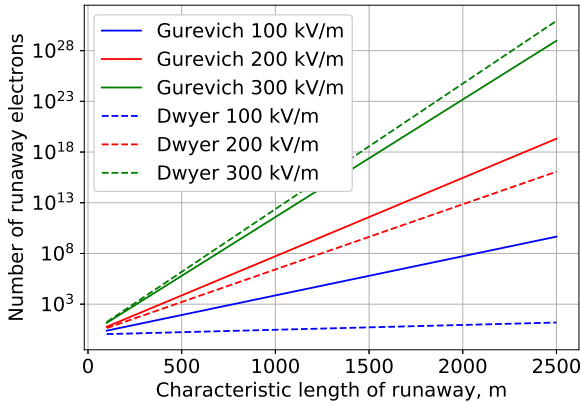}
		\caption{The number of runaway electrons in a single avalanche with respect to the avalanche length.}
		\label{fig:gur}
	\end{figure}

	A simple analytical estimation of the number of runaway electrons can be obtained on the basis of the theory of runaway breakdown~\cite{Gurevich:2001}. Let us consider acceleration and avalanche multiplication of electrons in uniform electric field. According to~\cite{Gurevich:2001} number of runaway electrons in the RREA grows exponentially with coordinate z:
	\begin{equation}
		\label{eq:exp}
		N(z) = N_0 \cdot e^{\frac{z}{l_a}}
	\end{equation}
	where $l_a$ --- characteristic length of generation of runaway electron, which can be estimated by the following formula:
	\begin{equation}
		l_a = a\frac{2 m c^{2}}{e} \frac{1}{E}
	\end{equation}
	Here $m$ --- electron mass, $c$ --- speed of light, constant $a \approx 11$, $E$ --- electric field value. The proposed model lead to an upper bound of $l_a$, because it does not take into account radiative losses, which is significant for higher electron energies (over $80~\un{MeV}$).
	Alternatively, one can use the empirical formula from Monte-Carlo simulation performed by Dwyer~\cite{Dwyer2007}:
	\begin{equation}
		\label{eq:dwyer}
		l_a = \frac{7300\un{kV}}{E - 276\un{ \frac{kV}{m}} \cdot \frac{n}{n_0}},
	\end{equation}
	where $E$ is the value of the electric field, $n$ - air concentration and $n_0$ - air concentration under normal conditions. Using either of this formula, we can calculate the number of runaway electrons depending on the size of the region with field and the field strength according to Eq.~\ref{eq:exp}. 
	Figure~\ref{fig:gur} shows the number of runaway electrons for different atmospheric conditions. For reasonable conditions  (region of vertical size $1200\un{m}$ with the electric field up to $200\un{kV/m}$) the number of runaway electrons does not exceed $10^{10}$.
	
	\section{Simulation with GEANT4}
	\label{sec:swg}
	A more accurate estimation of characteristic length of runaway electron avalanche could be obtained using GEANT4 transport code~\cite{Geant2003,Geant2006, Geant2016}, a standard tool for Monte-Carlo simulation in high energy physics (in this work we use version 4.10.05). The particle propagation mechanism used in GEANT4 is quite different from the one used in \cite{Oreshkin_2018}. GEANT4 uses particle tracks describing trajectories each particle and its descendants. The simulation in \cite{Oreshkin_2018} uses particle population evolution in time. The evolutionary approach is useful for low-energy particle and plasma dynamics but usually is not used for precise simulation of high energy particles. GEANT4 allows to track gamma-rays and positrons as well as electrons, which is important for electron energies above $10\un{MeV}$. To investigate the role of processes involving positrons and gammas, we conducted three types of simulations: 
	\begin{enumerate}
		\item Tracing electrons (and not considering any influence from gamma and positrons);
		\item Tracing electrons and gammas;
		\item Electrons, gammas and positrons are taken into account.
	\end{enumerate}
	
	All our simulations use a lower energy threshold for particle generation of $0.05\un{MeV}$ (particles born with lower energies are recorded, but do not propagate any further).
	
\begin{figure}[h]
		\centering
		\includegraphics[width=0.45\textwidth]{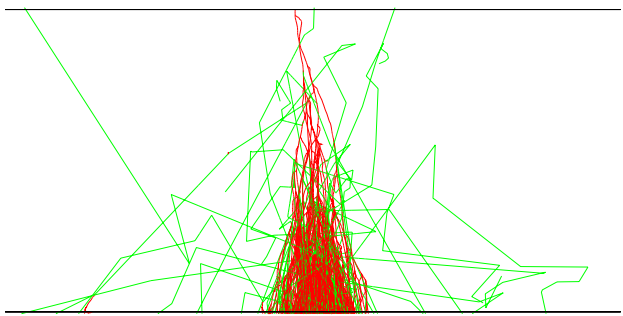}
		\caption{Example of GEANT4 simulation without positrons: red tracks --- electrons, green tracks --- gamma-rays.}
		\label{fig:geant4}
	\end{figure}

	Depending on the energy range, GEANT4 provides several models of physics processes, which are defined in physics lists. We use physics list \textit{G4EmStandardPhysics} , which takes into account the following electrons interactions:
	
	\begin{itemize}
		\item ionization loss,
		\item Coulomb scattering,
		\item multiple scattering,
		\item bremsstrahlung,
		\item gamma scattering and absorption,
		\item positron generation and annihilation.
	\end{itemize}
	It should be noted, that there is a physics list \textit{G4EmStandardPhysics\_opt4}. It contains the same physical processes but uses an advanced algorithm for tracking and requires much more processing time. Previous research shows that the basic physics list gives a higher estimate for the number of runaway electrons, so it could be used for upper boundary estimation~\cite{npmdwyer}.
	In the simulation, we use vertical cylinder cloud geometry. The radius of the cylinder, in this case, is defined to be much larger than cylinder height so the horizontal dimension is effectively infinite. The electric field is directed along the cylinder axis. An example of avalanche simulation without positrons is shown in Fig.~\ref{fig:geant4}.
	One of the main reasons behind the simulation approach in \cite{Oreshkin_2018} is to limit otherwise very large simulation time and memory footprint. GEANT4 in general shares the same problem, but allows us to set an upper limit by calculating only multiplication length in Gurevich model. For longer acceleration volumes, the effective number of produced electrons is smaller than for small ones possibly because higher energy electrons tend to lose energy by producing high energy bremsstrahlung photons rather than secondary electrons. Photons also could produce electron avalanches, but those avalanches are in general separated from the main avalanche by hundreds of meters.
	
	The simulation was carried out for following parameters of the system:
	\begin{itemize}
		\item air density $0.4\un{kg/m^3}$ (corresponds to atmospheric pressure $\sim 0.25\un{atm}$ or the height $10\un{km}$ for normal conditions);
		\item electric field --- $200\un{kV/m}$ (the maximum electric field typically measured in thunderclouds~\cite{rakov_uman});
		\item acceleration cell height --- $800\un{m}$ for simulation without positrons and $700\un{m}$ for simulation with positrons (as we will see in result this length is approximately equal to 10 characteristic lengths and allows to accept the estimation as consistent).
	\end{itemize}
	
\begin{figure}[h]
		\centering
		\includegraphics[width=0.45\textwidth]{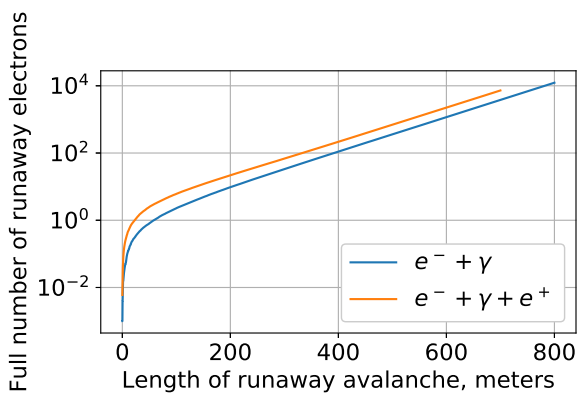}
		\caption{Full number of runaway electron depending on the length of runaway avalanche: blue line --- simulation without positrons, orange line --- with positrons.}
		\label{fig:sim}
	\end{figure}

	In each case, we inject 1000 initial electrons to the top of the cylinder (all results are normalized by the number of initial electrons). Results of simulation are shown in Fig.~\ref{fig:sim}. The introduction of gamma does not significantly affect the results, so we do not present results with electrons only and with electron and gamma separately.
	
	The simulation data was fitted by exponential function from Eq.~\ref{eq:exp} with $N_0 = 1$ and $l_a$ as a free parameter. The resulting value is $l_a \approx 85\un{m}$ without positrons, and $l_a \approx 78\un{m}$ with positrons. For comparison, Eq.~\ref{eq:dwyer}~(Dwyer's formula) predicts $l_a \approx 64\un{m}$ . Differences between our and Dwyer's result can be explained by the fact that Dwyer conducted simulation for normal condition and extrapolated results for another pressure. In addition, differences in the configuration of the Monte-Carlo model can contribute. Using received values for characteristic length, we can estimate the full number of runaway electrons in avalanches for different acceleration lengths. The results are presented in Table~\ref{tab:approx}. For reasonable acceleration cell lengths  $1200 - 1700\un{meters}$ it gives only $10^6-10^8$ runaway electrons, which is by many order less than $10^{17}$ - $10^{18}$ predicted by \cite{Oreshkin_2018}. $10^{17}$ - $10^{18}$~runaway electrons could be obtained only for cell lengths closer to $4000\un{meters}$ which are not possible in real atmospheric conditions~\cite[p.69-85]{rakov_uman}. It should be noted that we considered an avalanche created by a single seed electron, while cosmic rays are usually assumed to be a source of seed particles. But the flux of electrons of the required energies lies in the range of $10^3-10^4\un{m^{-2}s^{-1}}$~\cite{dwyer2003fundamental, sato2015analytical, sato2016analytical, dorman2013cosmic}. And taking into account that the movement of one avalanche through a cloud $1-1.5\un{km}$ length takes less than $10\un{\mu s}$ of the avalanche created by these electrons is quite separated in time and space. So that even if we take one single avalanche and a set of avalanches created with cosmic rays, then we still will not be able to achieve the value obtained by \cite{Oreshkin_2018}.
	
	\begin{table}[h]
		\centering
		\begin{tabular}{crrr}
			\hline
			& & \multicolumn{2}{r}{Number of runaway electrons} \\
			&   Length, m &   without positrons &  with positrons \\
			\hline
			\multirow{4}*{\rotatebox[origin=c]{90}{ simulation}} & 300 &  34.3  &  46 \\
			& 500 &  361     &  589 \\
			& 700 &  3802     &  7539 \\
			& 800 &  12350 &  --- \\
			\hline
			\multirow{5}*{\rotatebox[origin=c]{90}{extrapolation}}& 1200 &  1.4e+06 &  4.3e+06 \\
			& 1700 &  5.0e+08 &  2.5e+09 \\
			& 2000 &  1.7e+10 &  1.2e+11 \\
			& 4000 &  2.9e+20 &  1.3e+22 \\
			& 5000 &  3.7e+25 &  4.5e+27 \\
			\hline
		\end{tabular}
		\caption{Estimation of the total number of runaway electrons based on $700-800\un{meter}$ simulation. The first part of the table is taken from the simulation, the second part is the extrapolation of simulation data.}
		\label{tab:approx}
	\end{table}
	
	When discussing lightning breakdown, it is important to compute not only number of breakdown electrons, but also the total number of electrons and ions generated. This number could be very roughly estimated from energy conservation law. 
	
	The number of ions produced by a single RREA can be estimated using the following logic. Firstly, all of the energy runaway electrons receive from the electric field is spent on ionization, on the production of new runaway electrons and on acceleration. Received energy in a segment $[z, z + dz]$ is approximately $N(z) e E d z$. Here $e$ is the electron electric charge, and all of the runaway electrons are considered to move parallel to the $z$ axis. Further, the energy, spent on runaway electron production is about $\frac{d N(z)}{d z} \varepsilon_{0} d z$, where $\varepsilon_{0}$ is the mean kinetic energy of a runaway electron for conditions under consideration. Substitution of $\varepsilon_0$ in this formula also takes into consideration the acceleration of runaway electrons. According to Babich \cite{Babich2001}, the mean electron energy of RREA can be estimated as $\varepsilon_0 = e(E - E_c) l_a$, where $E_c$ is the critical electric field value. Consequently, in the frame of Dwyer's e-folding length, the mean energy is simply derived as $\varepsilon_0 = 7.3\un{MeV}$~\cite{Dwyer2007}. Finally, the number of ions is calculated as the subtraction of received energy and energy wasted on runaway electron production, all divided by ionization potential. Ionization potential is considered to be $15\un{eV}$. Therefore, after integration the following estimation of the number of ions is derived:
	\begin{equation}
		N_{ions}(z) = \frac{e E l_a - \varepsilon_0}{I} \left( e^{\frac{z}{\l_a}} - 1\right).
	\end{equation}
	For $z = 1000\un{m}$ we get $N_{ions} \approx 10^7$.
	
	\section{Discussion}
\begin{figure}[h]
		\centering
		\includegraphics[width=0.45\textwidth]{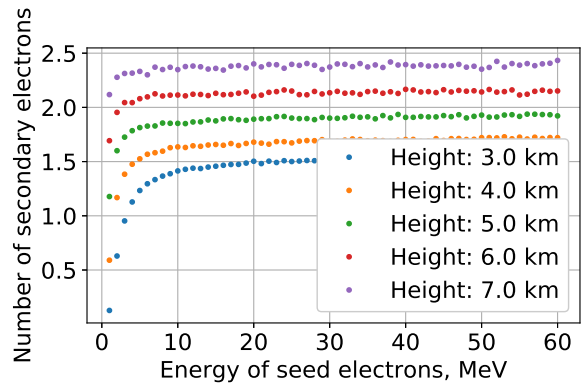}
		\caption{The number of electrons escaping $100~\un{m}$ length volume depending on the initial energy of the electron. Electric field is $200\un{kV/m}$. The figure shows that the number of secondary particles does not depend on initial particle energy for energy over $10\un{MeV}$}
		\label{fig:sec}
	\end{figure}

	Although the full Monte-Carlo simulation could not be carried out for a cell with a vertical size of more than $1000\un{m}$, the avalanche evolution on a larger scale could be characterized on the basis of extrapolation of the modeling results, since no physical mechanism that predicts a dramatic increase in the number of produced electrons for larger cells. 
	An increase in the cell length leads to growth of the maximum electron energy in the avalanche. But radiative losses prevent electrons from acquiring energies significantly higher than $60\un{MeV}$. Fig.~\ref{fig:sec} shows the number of secondary electrons depending on the initial electron energy. It could be seen that there is no significant yield change for initial energies above $5-10\un{MeV}$. Growth of radiative losses lead to increase of number secondary gamma-rays, which could produce secondary electrons, but the following two circumstances should be kept in mind:
		\begin{itemize}
			\item generation of runaway electrons by gammas can lead to the increase of number of runaway electrons by several times, not several orders of magnitude,
			\item high-energy gammas in air are characterised by mean free path (MFP) comparable with characteristic cloud size (MFP of $1\un{MeV}$ photons is about $380\un{meters}$ at the altitude $10\un{km}$~\cite{xcom} and increases with raise of photons energy), for this reason secondary electrons are more likely generated by these gammas outside of the region of localisation of the initial avalanche; and, consequently, the question of whether secondary electrons belong to the initial avalanche is controversial: these electrons create only a subtle change of electron density in the region of the initial avalanche.
	\end{itemize}
	
	Dwyer positron feedback mechanism \cite{Dwyer2012feedback} could, in theory, produce a significant increase in the number of runaway electrons. Dwyer's calculation independently verified in work~\cite{skeltved2014}, but the recent research shows that the feedback value in this mechanism was rather overestimated~\cite{npmdwyer}.
	
	In any case, both in this article and in \cite{Oreshkin_2018} we consider only forward avalanche motion without any feedback mechanisms. Still, our results show that it is not correct to completely ignore the positrons, which are more significant for a cell with a vertical size exceeding $1000\un{m}$ . The increase in the region in which acceleration occurs leads to an increase in the number of high-energy photons for which the production of pairs makes the important contribution to the cross-section for interaction with matter~\cite{heitler1984quantum, Geant2016} (In air, cross-section of pair production exceeds cross-section of other process for photons with energy over $26~\un{MeV}$~\cite{xcom}).
	
\begin{figure}[h]
		\centering
		\includegraphics[width=0.45\textwidth]{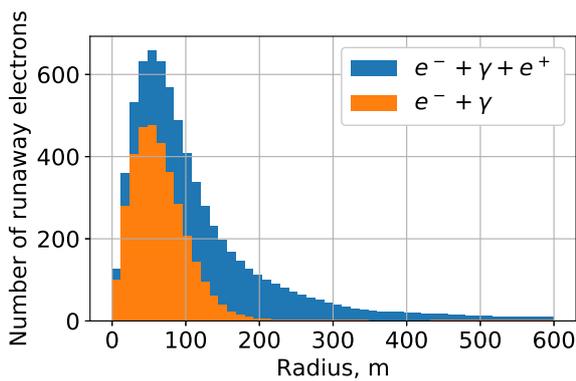}
		\caption{The distribution of runaway electrons horizontal distance from initial electron creation point for avalanches with length $700\un{meters}$ (electric field is $200\un{kV/m}$): orange --- without positrons, blue --- with positrons. }
		\label{fig:rad}
	\end{figure}

	The radial distribution of runaway electrons production is demonstrated in Fig.~\ref{fig:rad}. It could be seen that electrons have a wide horizontal distribution. This fact further reduces the credibility of claims that those avalanches are the source of lightning breakdown since the produced charge is distributed across rather a large region. The horizontal spreading is significantly larger when positrons are taken into account because in this case, photons could transfer the secondary avalanches far from initial ones.
	
	At the end of the discussion, we would like to note that in paper~\cite{Oreshkin_2018} and in our works the field in the cloud was considered to be uniform, whereas there is a debatable question about the existence of areas in the cloud that are not accessible for measurement and at the same time contain large fields~\cite{dwyer2003fundamental}. Influence of these areas depends on runaway electron generation length in them and may be subject to particular research.
	
	\section{Conclusions}
	
	In this work, we calculated the number of runaway electrons in conditions expected in real thunderclouds (size of the strong field region up to $1200\un{m}$ and field strength up to $200\un{kV/m}$). The simulation itself covers only cell lengths up to 800~m, but it is clear that electron production could be extrapolated. The simulations provide an upper bound for the number of runaway electrons of about $10^6$ runaway electrons or $10^{10}$ total ionization per one seed electron for length about $1200\un{meter}$, which is in accordance with \cite{Gurevich:2001, Dwyer2013_radio} and contradicts to the value presented in \cite{Oreshkin_2018} (more than $10^{16}$ runaway electrons). Especially since Oreshkin’s report considers only electrons, and our results show that the introduction of gamma and positron generation increases the yield of runaway electrons.
	Still, even with those interactions, the resulting numbers are by several orders lower than the ones shown in \cite{Oreshkin_2018}. Although the origin of the discrepancy could be investigated further, it is clear that the non-relativistic particle dynamics model could not be used to describe the generation of enough ionization to produce a lightning discharge.
	
	\acknowledgments
	This work is supported by the Russian Science Foundation under grant No. 17-12-01439.
	
	\bibliographystyle{eplbib.bst}
	\bibliography{references}
\end{document}